# Strong interaction of slow electrons with near-field light visited from first principles

**Nahid Talebi**

Institute for Experimental and Applied Physics, Christian Albrechts University, Leibnizstr. 19, 24118 Kiel, Germany

E-mail: talebi@physik.uni-kiel.de

**Abstract-** Strong interaction between light and matter waves, such as electron beams in electron microscopes, has recently emerged as a new tool for understanding entanglement. Here, we systematically investigate electron-light interactions from first principles. We show that enhanced coupling can be achieved for systems involving slow electron wavepackets interacting with plasmonic nanoparticles, due to simultaneous classical recoil and quantum mechanical photon absorption and emission processes. For slow electrons with longitudinal broadenings longer than the dimensions of nanoparticles, phase-matching between slow electrons and plasmonic oscillations is manifested as an additional degree of freedom to control the strength of coupling. Our findings pave the way towards a systematic and realistic understanding of electron-light interactions beyond adiabatic approximations, and lay the ground for realization of entangled electron-photon systems and Boson-sampling devices involving light and matter waves.

Strong light-matter interaction has been intensively addressed from various physical aspects. In condensed matter, strong coupling leads to the emergence of new quasiparticles known as polaritons [1]. Within solid state systems, strong Coulomb interactions leads to quantum mechanical correlations and entangled systems, being manifested in superconducting states of matter [2]. Photon excitations have been proposed as new degrees of freedom for controlling correlations [3]. In other words, photoinduced correlations appear as novel tools for controlling energy, spin, and lattice degrees of freedom. However, the complexity of systems involved, i.e., the combination of solid-state lattice degrees of freedom with many body interactions and photon excitations, hinders a thorough and systematic investigation of emerging phenomenon. Therefore, systems with less complexity could serve for better understanding of entanglements and correlations in matter waves [4].

Photon-induced near-field electron microscopy (PINEM) has been proposed in 2008 as a tool for investigating ultrafast dynamics of physical and chemical reactions with electron probes [5,6]. In PINEM, synchronous electron and photon pulses are employed, within a pump-probe excitation platform, to coherently derive photoinduced polarizations in the sample with laser pulses and to simultaneously probe them with electron pulses (Fig. 1a). Within weak interaction regimes, PINEM, or more specifically electron energy-gain spectroscopy [7], is suited to probe the dynamics of material excitations such as plasmons [8-14], and holographic imaging of near-field dstributions [15]. Multiple photon absorption and emission processes are encountered and therefore, PINEM spectroscopy typically results in multiple resonances, i.e., a frequency comb, displaced by the photon excitation energy, with a prominent zero-loss peak [16]. Within the weak interaction regime, frequency combs demonstrate a monotonic decreasing intensity versus their excitation energies (see Fig. 1b and c). Depending on the quantum mechanical fluctuations and the statistics of photonic excitations, either symmetric or asymmetric spectral features might be observed [17].

Strong interactions between electron pulses and photon excitations in contrast, provides ways for coherently driving non-equilibrium interfering motions within electron wavepackets leading to emerging novel phenomena. Strong interactions might ultimately lead to entanglement between electrons and photons, as theoretically proposed by O. Kfir [18]. It was proposed that phase-matching between electron and photon excitations, fostered by means of whispering gallery modes, can result in an enhanced coupling strength. Strong interaction lead to a partial depletion of energy from the zero-loss peak and generation of higher order elastic and inelastic photon processes (Fig. 1c). Most recently, experiments demonstrated that such strong interactions are achievable by using synchronous electron and photon excitations in a whispering gallery resonator [17] or in along an interface [19]. However, theories involved in describing such strong-interaction phenomenon are still within adiabatic approximations, neglecting the role of diffraction and experienced electron recoils. Therefore, a systematic investigation of electron-light interactions and the parameters involved is still incomplete. A thorough investigation from first-principle will pave the way towards a better understanding of correlations between photon and electron excitations.

Here, using first-principle analysis, beyond routinely employed perturbation approximations and 1-dimensional models, we systematically investigate the interaction of slow and fast electron wavepackets with near-field plasmonic excitations. We define the angle-resolved differential energy expectation value as a quantity for distinguishing unique features than can be observed as a result of strong interaction. We therefore, are able to study both elastic and inelastic processes and configure photoinduced classical and quantum paths leading to weak and strong interactions, as well as diffraction and attosecond bunching. We also show that the spacing between the attosecond bunches can be controlled by the size of the nanostructure, particularly in the case of synchronous electron motion and near-field oscillations. Moreover, we propose and study systems that can foster enhanced coupling strengths. It is further demonstrated that strong coupling between slow and fast electron excitations lead to a simultaneous modulations in phases and amplitudes of the electron wavepackets. The strength of coupling can be hence characterized as the visibility of spectral interference features as well as energy-split value, where the latter is caused due to the classical wiggling motion of the electron in photoinduced near-field zone.

We consider the interaction of a moving electron wavepacket with photoinduced ultrafast charge oscillations inside a gold nanorod (see Fig. 1a). In all cases discussed here, the time delay between electron and laser pulses are set to zero, thus we intend to discuss other parameters involved in the strength of electron-light interactions. The parameters underlying the strength of the interaction; in other words the coupling strength, are center group velocity ($v_{el}$), longitudinal ($W_L$), and transverse ($W_T$) broadenings of the electron wavepacket, as well as laser excitation parameters such as its peak electric-field amplitude, polarization, wavelength, and spectrum. We assume that our electron and photon excitations are mutually coherent and their statistics are manifested by coherent states; therefore are classically describable. We assume an initially Gaussian electron wavepacket with $W_x = 20\,\text{nm}$ interacting with the sample, which is composed of a gold nanorod excited by a linearly polarized laser pulse along the $x$ direction with the center wavelength of $\lambda = 800\,\text{nm}$, temporal broadening of 16 fs, and its direction of propagation being perpendicular to the electron trajectory. We employ a recently developed time-dependent numerical toolbox [20,21] to investigate the dynamics of the interaction of electron wavepackets with the laser induced plasmonic oscillations in the gold nanorod. In this approach Maxwell's equation and Schrödinger equations are combined in a time-dependent loop, using the minimal coupling Hamiltonian as

$$-\left(\hbar^2/2m_0\right)\nabla^2\psi(\vec{r},t)-\left(i\hbar e/m_0\right)\vec{A}(\vec{r},t)\cdot\vec{\nabla}\psi(\vec{r},t)+\left(e^2/2m_0\right)\left|\vec{A}(\vec{r},t)\right|^2\psi(\vec{r},t)-e\varphi(\vec{r},t)\psi(\vec{r},t)=i\hbar\dot{\psi}(\vec{r},t)$$

, where the Coulomb gauge has been implied and $\varphi$ is the scalar potential, $\vec{A}$ the vector potential, and $\psi$ the single-electron wavepacket. Maxwell and Schrödinger equations are solved using finite dfference tme domain technique and pseudospectral approaches, respectively. The required parameters like the potentials are mapped between Maxwell and Schrödinger simulation domains using inter- and extrapolation techniques. The sample is modeled by its dispersive dielectric function. We calculate the scalar and vector potentials from time-dependent field components using Fourier transformation (See supplemental section S1). Dynamics of the interaction and the ultrafast modulations in the electron wavepackets are thus captured within the space-time domain. Therefore, the electron wavepacket after the interaction contains the required information to interpret the outcomes of spectroscopy and diffraction measurements.

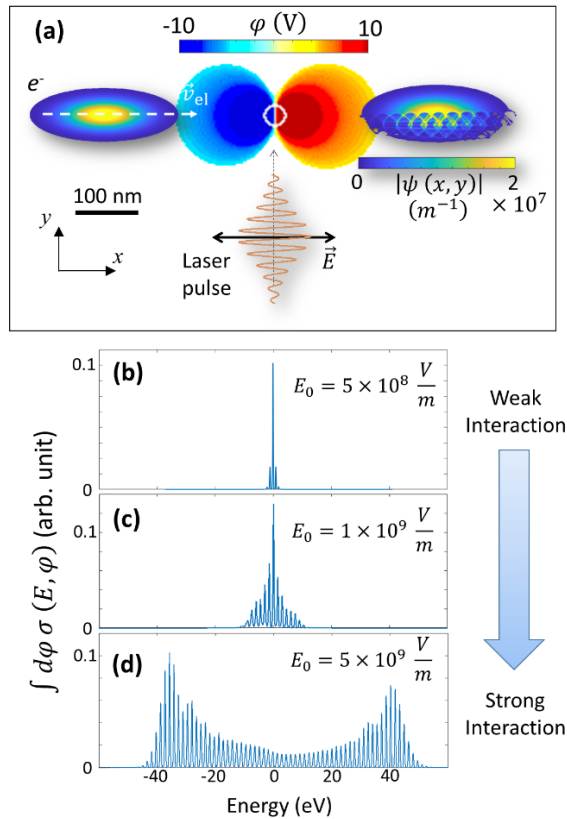

**FIG. 1. Weak and strong electron-light interactions.** (a) Interaction of an electron wavepacket with laser-induced dipolar plasmon oscillations in a gold nanorod with the radius of 15 nm. Demonstrated is the amplitude of the electron wave function before and after the interaction, and the electromagnetic scalar potential. Electron has the kinetic energy of 200 eV, and its initial longitudinal and transverse broadenings are 20 nm and 2nm, respectively. Laser field amplitude, wavelength, and temporal broadening are $2 \times 10^9$ V m$^{-1}$, 800 nm, and 16 fs, respectively. PINEM spectra for an electron wavepacket with the same specifications as above, but a different kinetic energy of 400 eV, interacting with the same system as stated above, excited at different laser field amplitudes of (b) $E_0 = 5 \times 10^8$ V m$^{-1}$, (c) $E_0 = 1 \times 10^9$ V m$^{-1}$, and (d) $E_0 = 5 \times 10^9$ V m$^{-1}$, specifying an increasing coupling strength and therefore photon absorption and emission orders.

Electron spectroscopy detectors such as a magnetic prism spectrometer detect the kinetic energy of electrons and map the energy distribution over a charge coupled diode array. The expectation value of the kinetic energy operator is the measureable physical quantity, given by

$$\langle \psi(x,y;t\to\infty)|\hat{H}_K|\psi(x,y,t\to\infty)\rangle = \frac{\hbar^2}{2m_0}\iint dk_x dk_y \left(k_x^2+k_y^2\right)|\tilde{\psi}(k_x,k_y,t\to\infty)|^2 \quad (1)$$

where $(x,y)$ and $(k_x,k_y)$ denote the real space and reciprocal space coordination, and $\tilde{\psi}$ is the Fourier transform of the wavefunction. By $t\to\infty$ we imply a given time after the interaction, where the spectroscopy measurement is performed. Eq. (1) is recast in the cylindrical coordination as $\iint d\varphi dk_r \left(\hbar^2 k_r^2/2m_0\right) k_r |\tilde{\psi}(k_r,\varphi,t)|^2$. Angle-resolved differential energy expectation value is therefore represented as

$$\sigma(E,\varphi)=\frac{d}{dEd\varphi}\langle\psi(x,y,t)|\hat{H}|\psi(x,y,t)\rangle = \frac{m_0}{\hbar^2}E|\tilde{\psi}(E,\varphi,t)|^2 \quad (2)$$

with $\varphi=\tan^{-1}(k_y/k_x)$ and $E=\hbar^2 k_r^2/2m_0$. Eq. (2) is referred to as the angle resolved PINEM spectrum. The overall PINEM spectrum is given by an integration of the form $\Sigma(E)=\int_{-\varphi_0}^{+\varphi_0}d\varphi\,\sigma(E,\varphi)$, where the span of angular integration is given by the spectrometer acceptance angle.

We first consider the interaction of a slow electron wavepacket at the energy of 400 eV with a gold nanorod with the radius of $r=15\,\text{nm}$. Other parameters for laser excitation and the electron wavepacket are given above. An obvious parameter for controlling the coupling and hence the interaction strength is the laser peak electric-field amplitude. By systematically increasing the field amplitude, different domains from weak interaction to strong interaction are spanned (see Fig. 1b to d). In general, an energy comb is observed, where the spacing between the peaks are exactly the photon energy $\hbar\omega_{ph}=1.55\,\text{eV}$, demonstrating multiple photon absorption and emission peaks in a kinematic or dynamic process [22]. At lower field amplitudes below $E_0=2\times10^9\,\text{V m}^{-1}$ and within the weak interaction regime, the resonance amplitudes monotonically decrease versus their orders. By increasing the field amplitude to $E_0=5\times10^9\,\text{V m}^{-1}$, higher photon absorption and emission processes up to 30 orders are observed. Moreover, the highest amplitude resonance is not located at $E=0$ any more, but at $E=-37.2\,\text{eV}$ and $E=+43.4\,\text{eV}$, and the spectrum is also not symmetrically distributed around the origin.

Besides the field amplitude, other parameters can be used to control the coupling strength. Keeping the field amplitude as high as $E_0=5\times10^9\,\text{V m}^{-1}$, we change the nanoparticle geometry to an elliptical form with $r_x=75\,\text{nm}$ and $r_y=15\,\text{nm}$, denoting the radius of the ellipse along x and y directions, respectively. Moreover, the temporal broadening of the laser is increased to 21 fs. Only photon absorption and emission peaks up to $\pm 10$ eV are observed along the $\varphi=0$ axis, and the angular span of the final wavepacket remains within $\delta\varphi=3°$ at $E=0$ and $\delta\varphi=1°$ at higher energy loss and gain regions (FWHM values are mentioned) (Fig. 2a).

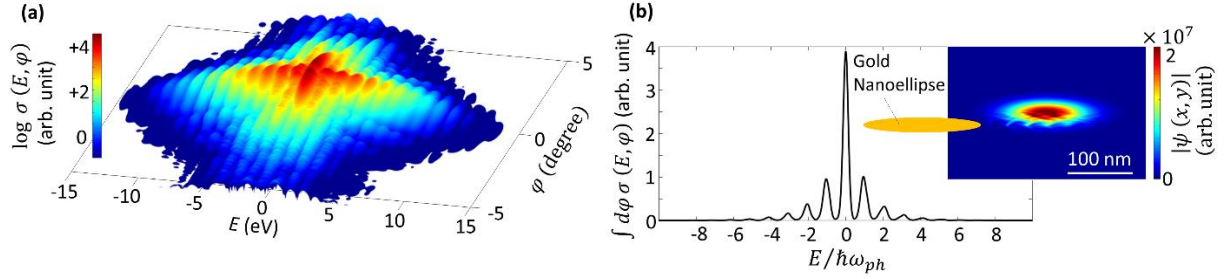

**FIG. 2.** Weak interactions between an electron and photon-induced plasmon oscillations in a gold nanorod with an *elliptical* cross section. The electron has a kinetic energy of 400 eV and initial longitudinal and transverse broadenings of 20 nm and 2 nm, respectively. Laser field-amplitude, wavelength, and temporal broadening are $5 \times 10^9$ V m$^{-1}$, 800 nm, and 16 fs, respectively. (a) Angle-resolved differential expectation value of energy and (b) PINEM spectrum. Inset demonstrates the amplitude of the electron wavepacket after its interaction with the system.

Note that the traveling time of the center of the electron wavepacket through the near-field region is approximately 12 fs, which is less than the broadening of the laser pulse. Therefore, for all calculations, the electron experiences an almost constant field amplitude. Despite an increase in the interaction time, the overall coupling strength is less than the case of interaction with a circular nanoparticle with the radius of $r = 15$ nm. Therefore, the size of the nanoparticle plays an important role in controlling the strength of the interaction.

The coupling constant for inelastic PINEM interactions along the electron center of mass trajectory is specified by the parameter $g = (e/2\hbar\omega)\int dk_y \, \tilde{E}_x(k_x = \omega_{\text{ph}}/v_{\text{el}}, k_y; \omega_{\text{ph}})$ [11], where $\tilde{E}_x$ is the Fourier transform of the $x$-component of the electric field, and $k_{x,c} = \omega_{\text{ph}}/v_{\text{el}}$ implies the momentum selection rule specifying the role of near-field interactions. In other words, only near-field photons with the specified wavenumber can derive inelastic transitions in the electron wavepackets along the $y = 0$ ($\varphi = 0$) axis. Particularly for an electron at the kinetic energy of 400 eV ($v_{\text{el}} = 0.0395c$), $k_{x,c} = 25k_0$, where $k_0$ is the photon wavenumber in free space. Therefore, only nanoparticles with dimensions much smaller than the free space wavelength of $\lambda_0$ can provide enough momentum for inelastic electron-light interactions. Therefore, for nanoparticles with the diameter of $2r = 30$ nm, being approximately 26 times smaller than $\lambda_0$, the required wavenumber for inelastic interaction can be provided, whereas the elliptical structure, with $2r_x = 150$ nm, the required wavenumber is hardly achieved. PINEM spectrum calculated for elliptical structure demonstrates that much less photon absorption and emission orders are occupied (Fig. 2b), compared with the spectrum shown in Fig. 1d. Therefore, regardless of the increase in both interaction length and time, the overall coupling constant and therefore the interaction strength remains weak.

In addition to inelastic photon absorption and emission channels, significant elastic photoinduced diffraction channels are also observed. For the case of strong interactions, time-dependent quantum coherent transitions constructively interfere over the time, leading to significant ultrafast modulations in both the amplitude and the phase of the electron wavepacket (See Fig. 3a and b and Supplemental Movies). The overall interaction of the electron wavepacket with the broadening of $W_{\text{L}} = 20$ nm with the nanorod

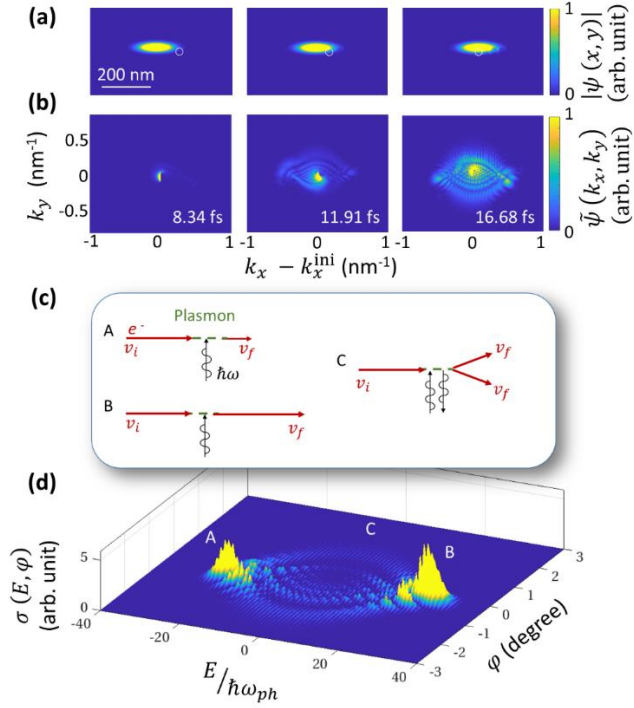

**FIG. 3.** Strong Interaction for an electron at the kinetic energy of U=200 eV, interacting with the laser-induced localized plasmon oscillations in a gold nanorod with the radius of 15 nm. Electron has the kinetic energy of 400 eV, and initial broadenings of 2nm and 20nm at the transverse and longitudinal directions. The laser field amplitude is $5 \times 10^9\ V\ m^{-1}$ and its wavelength and temporal broadening are 800 nm and 16 fs, respectively. Amplitude of the electron wavepacket at the (a) spatial and (b) momentum space. (c) Single (A, B) and two-photon (C) processes, leading to electron acceleration, deceleration, and diffraction, respectively. (d) Map of the angle-resolved differential expectation value of energy, demonstrating the effect of diffraction and energy-loss and gain processes on the electron wavepacket (See supplementary movies 1 and 2).

with the radius of $r = 15$ nm, takes place only 16 fs, comparable to the temporal broadening of the laser pulse. Moreover, in addition to longitudinal fine-structures resulting in an energy comb specified in Fig. 1d, elastic diffraction peaks are also observed (See Fig. 1b and Supplemental Sec. 2), similar to the Kapitza-Dirac effect taking place due to the interaction of the electron wavepacket with standing-wave light patterns in free-space [23,24] or surface plasmon polaritons [25]. Electron-photon interactions leading to diffraction are two-photon processes, unlike inelastic photon absorption and emission processes that are single-photon processes [26]. In other words, interactions leads to simultaneous absorption and stimulated emission processes, therefore the net energy exchange between photon and electrons remain negligible for two-photon processes [27]. However, dissimilar to the Kapitza-Dirac effect, for which the diffraction orders are distributed along an Ewald sphere [27], for the current near-field mediated diffraction phenomenon, the diffraction orders are positioned along distinguished eye-like $k_x - k_y$ surfaces. These processes are particularly distinguished in the calculated angle-resolved differential energy expectation value (see Fig. 3d). Along the $\varphi = 0$ axis, pronounced energy loss and gain peaks at the orders of $-24\ \hbar\omega_{\text{ph}}$ and $+28\ \hbar\omega_{\text{ph}}$ as well as fine spectral features at the order of photon energies are observed. In contrast, diffraction peaks along $E = 0$ axis are displaced by orders specified by $\delta k_y^{\text{el}} = 2\left|\kappa_y^{\text{ph}}\right| = 2\sqrt{k_{x,c}^2 - k_0^2} = 2k_0/\beta\gamma$, where

$\beta = v_{el}/c$ and $\gamma$ is the Lorentz factor. $\kappa_y^{ph}$ is the damping ratio for evanescent tail of plasmons along the y-axis. For slow electrons, $\delta k_y^{el}$ is significantly larger than the Kapitza-Dirac diffraction orders of $2k_0$ and therefore can be easily retrieved. The overall arrangement of diffraction orders at different energies though depends on both the electron velocity and the optical near-field momentum distributions and thus the topology of nanoparticles (See Supplemental Fig. S3).

Therefore, dimensional aspects are crucial for achieving an enhanced coupling constant, both due to the requirements for momentum matching, but also to foster synchronous motions between near-field oscillations and electron wavepackets. For slow electrons, the extend of the near-field zone responsible for inelastic interaction is tightly bound to the nanoparticle surface. Consequently, electron wavepackets are effectively shaped according to the geometry of the nanoparticles (Fig. 4). Particularly, when $T_{ph} = 2r/v_{el}$, with $T_{ph} = \lambda_0/c$, a synchronous motion is achieved, resulting in attosecond bunching of the electron wavepacket with interspacing between the bunches dictated by the size of the nanoparticle (Fig. 1a and supplementary Fig. S5). For a gold nanoparticle with the radius of 15 nm, the required electron kinetic energy for achieving synchronicity is $U = 359\,\text{eV}$. Therefore, an electron at this kinetic energy interacting with a gold nanorod with the radius of 15 nm is bunched into a series of ultrashort pulses (with the duration of 253 as), with an inter-distance spacing of 30 nm, corresponding to 2.53 fs temporal spacing (See

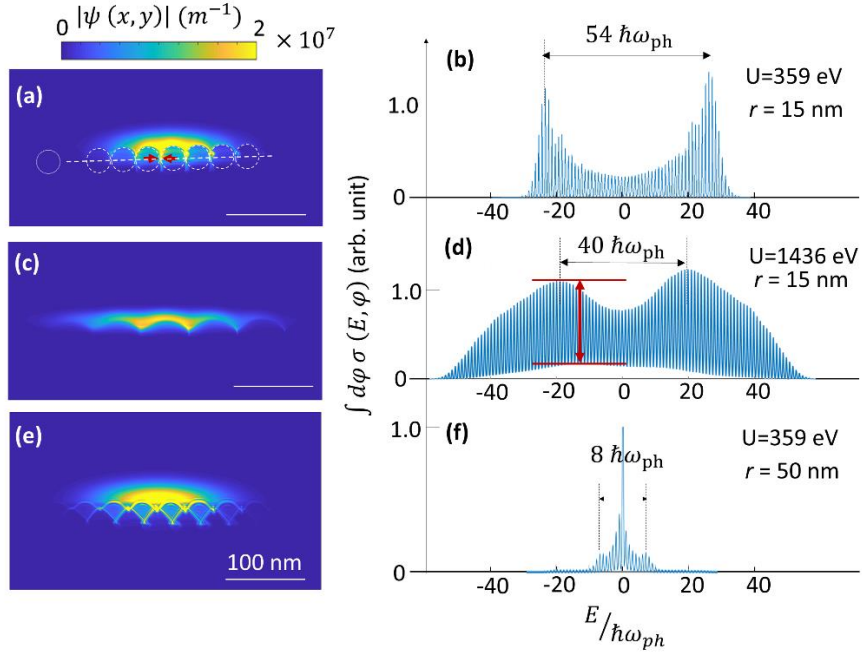

**FIG. 4.** Controlling coupling strength by means of phase-matching (synchronicity) between electron and light oscillations. (a, c, d) amplitude of the electron wavepacket and (c, d, f) PINEM spectra, after the interaction with a gold nanorod excited with an x-polarized laser pulse at the wavelength of 800 nm, field amplitude of $5 \times 10^9$ V m$^{-1}$, and temporal broadening of 16fs. (a, b) $U = 359$ eV and $r = 15$ nm, (c, d) $U = 1436$ eV and $r = 15$ nm, (e, f) $U = 359$ eV and $r = 50$ nm.

Supplemental Fig. S5). Within this configuration, the PINEM spectrum, integrated over the complete angular range, demonstrates an energy split at the order of $54\hbar\omega_{ph}$, and a fine-structure energy comb occupying up to 30 photon emission and absorption orders (Fig. 1b). Moreover, the visibility of the spectral interference fringes, defined by the ratio of the maximum to minimum values, approaches infinity, demonstrating the role of perfect synchronicity (mutual coherence) in achieving strong electron-photon interactions.

By increasing the kinetic energy of the electron to $U = 4 \times 359$ eV, corresponding to $U = 1436$ eV, less pronounced photon bunching is achieved (Fig. 1c). The energy split of the PINEM spectrum is reduced to $40\hbar\omega_{ph}$, and the visibility of the interference fringes is degraded. Indeed, by increasing the electron velocity to as high as 10 keV, the visibility of the interference fringes is even further reduced (See Supplemental Fig. S4 and Supplementary Note S4.). This is due to the fact that the synchronicity between electron and near-field photoinduced oscillations are no longer achievable. However, the span of the fine-structure energy comb is significantly increased to $\pm 48\hbar\omega_{ph}$, demonstrating the impact of momentum-matching criterion on electron-light interactions; hence, the condition $k_{x,c} = \omega_{ph}/v_{el}$ is much easier attained for fast electron wavepackets.

The strength of coupling can be as well altered by changing the size of nanoparticle. By keeping the kinetic energy of the electron at $U = 359$ eV, but changing the nanoparticle radius to 50 nm, a more complicated electron bunching effect is observed, highlighting the impact of multimodal near-field excitations and the race between their effects on modulating the shape of the electron wavepacket (Fig. 1e). The extend of the PINEM spectrum though is drastically reduced, reaching up to only $\pm 10\,\hbar\omega_{ph}$ spectral range (See Supplemental Fig. S3 for electron wavefunction representation in momentum space), meaning that the quantum interference paths originating from various modes do not lead to a constructive interference pattern.

Therefore, the overall shape of $\sigma(E,\varphi)$ is specified by the electron velocity, as well as structural topology and material excitation, since the latter affect the near-field wavenumbers and multipolar distributions (see supplemental section 2). $\sigma(E,\varphi)$ serves as an experimental observable with valuable information for recovering material excitations, additional to fundamental understanding of electron-photon interactions. For example, electron-holography using a point-projection electron microscopy setup can be used to acquire $\sigma(E,\varphi)$ function [28]. However, still a suitable inverse approach has to be developed and employed to facilitate retrieving information about structural excitations from $\sigma(E,\varphi)$.

Additionally, though not shown here, multi-electron pulses with more than a single electron per pulse, can be used in PINEM-like experimental setups to coherently derive many-body interactions and control photoinduced correlations between electron waves, serving as a better understanding of electron-electron correlations and entanglements.

In summary, our systematic analysis of near-field mediated electron-light interactions highlights the impact of structural parameters as well as mutual coherence and phase-matching on the ultrafast modulations leading to photoinduced constructive interference paths within the electron wavepacket. Using the developed first-principle numerical toolbox, the effect of strong electron-light interactions on both elastic and inelastic channels is investigated. It is proposed that both laser and electron wavepacket parameters underlie the strength of electron-light coupling, with phase matching appearing as an important control parameter. Furthermore, conclusive remarks on the impact of the current work on facilitating systematic understanding of many-body correlations are provided.

This project has received funding from the European Research Council (ERC) under the European Union's Horizon 2020 research and innovation programme, grant agreements No. 802130 (Kiel, NanoBeam),